\documentclass[a4paper]{jpconf}
\usepackage{graphicx}
\usepackage{gensymb}

\begin{document}

\newcommand{\tn}[1]{\textnormal{#1}}
\newcommand{\pPb}{\textnormal{p--Pb}}
\newcommand{\RpPb}{\ensuremath{R_\mathrm{pPb}}}
\newcommand{\PbPb}{\textnormal{Pb--Pb}}
\newcommand{\pp}{\ensuremath{\mbox{p}\mbox{p}}}
\newcommand{\snn}{\ensuremath{\sqrt{s_\tn{NN}}}}
\newcommand{\kt}{\ensuremath{k_\mathrm{T}}}\newcommand{\kT}{\kt}
\newcommand{\pT}{\ensuremath{p_\mathrm{T}}}
\newcommand{\ptch}{\ensuremath{p_\mathrm{T,\,ch.}}}
\newcommand{\GeV}{\textnormal{ GeV}}
\newcommand{\abs}[1]{\left| #1 \right|}

\title{Measurement of inclusive jet spectra in $\pp$, $\tn{\bf{p--Pb}}$, and $\tn{\bf{Pb--Pb}}$ collisions with the ALICE detector}

\author{R\"udiger Haake for the ALICE Collaboration}

\address{Institut f\"ur Kernphysik, Westf\"alische Wilhelms-Universit\"at M\"unster, 48149 M\"unster, Germany}

\ead{ruediger.haake@uni-muenster.de}

\begin{abstract}
Highly energetic jets are sensitive probes for the kinematic properties and the topology of high energy hadron collisions. Jets are collimated sprays of charged and neutral particles, which are produced in fragmentation of hard scattered partons from an early stage of the collision.
In ALICE, jets have been measured in $\pp$, $\pPb$, and $\PbPb$ collisions at several collision energies. While analyses of $\PbPb$ events unveil properties of the hot and dense medium formed in heavy-ion collisions, $\pp$ and $\pPb$ collisions can shed light on hadronization and cold nuclear matter effects in jet production. Additionally, $\pp$ and $\pPb$ serve as a baseline for disentangling hot and cold nuclear matter effects. A possible modification of the initial state is tested in $\pPb$ analyses.
For the extraction of a jet signal, the exact evaluation of the background from the underlying event is an especially important ingredient. Due to the different nature of underlying events, each collision system requires a different analysis technique for removing the effect of the background on the jet sample.
The focus of this publication is on the ALICE measurements of nuclear modification factors connecting $\pPb$ and $\PbPb$ events to $\pp$ collisions. Furthermore, the radial jet structure is explored by comparing jet spectra reconstructed with different resolution parameters.
\end{abstract}

\section{Introduction}

Jets can conceptually be described as the final state produced in a hard scattering. Therefore, jets are an excellent tool to access a very early stage of the collision. The jet constituents represent the final state remnants of the fragmented and hadronized partons that were scattered in the reaction. While all the detected particles have been created in a non-perturbative process (i.e. by hadronization), ideally, jets represent the kinematic properties of the originating partons. Thus, jets are mainly determined by perturbative processes due to the high momentum transfer and the cross sections can be calculated with pQCD. This conceptual definition is descriptive and very simple, the technical analysis of those objects is quite complicated though.\\

In pp collisions, jets can serve as a probe for undisturbed pQCD physics. In $\PbPb$ collisions, the medium effects like those from the quark-gluon plasma (QGP) can be probed. For $\pPb$ collisions, formation of a hot, dense, and extended medium is not expected and therefore cold nuclear matter effects, e.g. those described by nuclear parton distribution functions (nPDFs), should be dominant.\\

Here, the focus is on the inclusive jet results currently available from the ALICE experiment. Results and conclusions of recent ALICE measurements in pp, $\pPb$, and $\PbPb$ are presented in Sec.~\ref{sec:Results}. Beforehand, a short description of the ALICE detector with regard to jet measurements is given in Sec.~\ref{sec:ALICEDetector}. Additionally, important details of the analyses are shown in Sec.~\ref{sec:Techniques}.

\section{The ALICE detector}
\label{sec:ALICEDetector}

\begin{figure}[h]
\centering
\includegraphics[width=22pc]{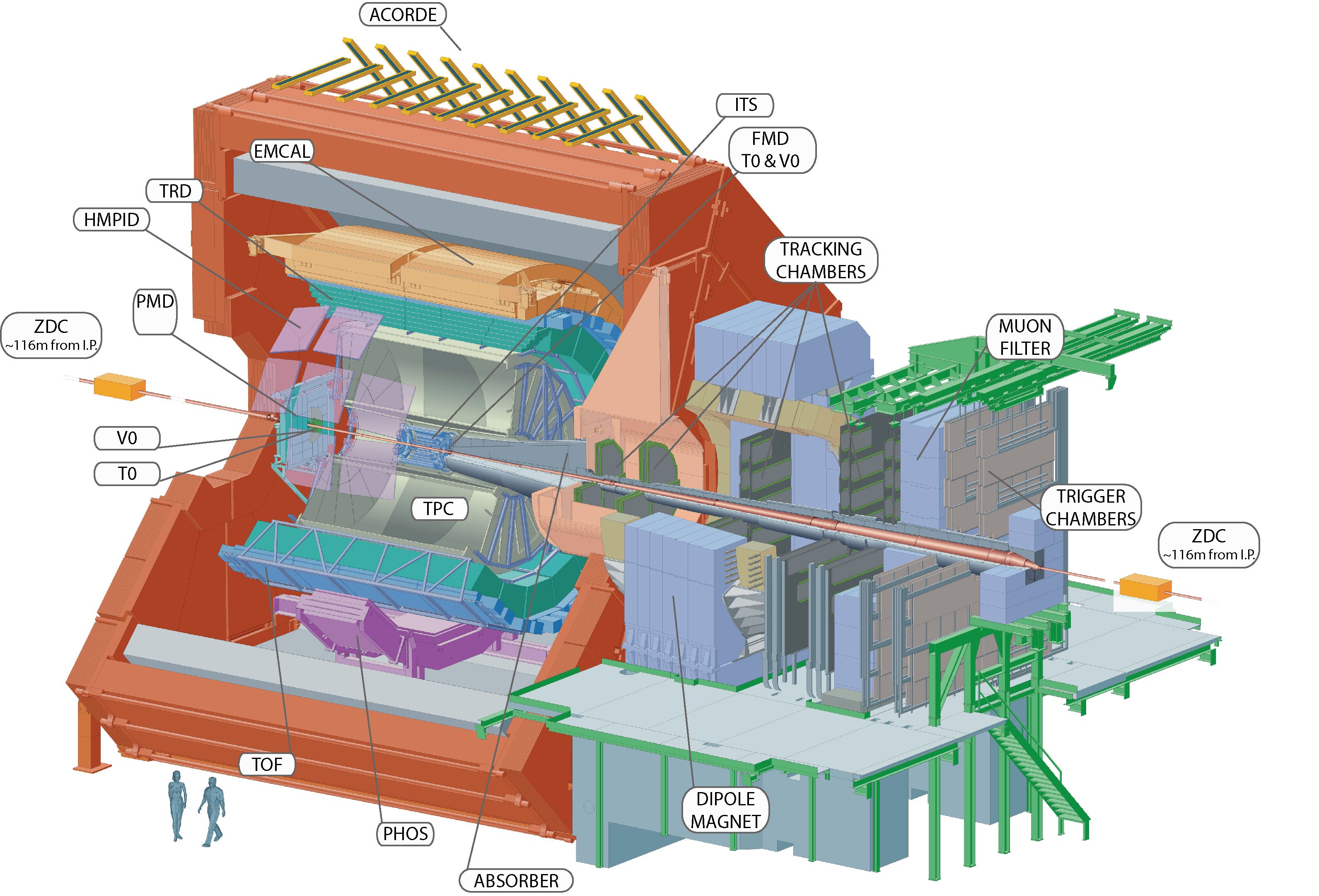}\hspace{2pc}%
\caption{\label{fig:ALICE_detector}A sketch of the ALICE detector}
\end{figure}

The ALICE detector (see Fig.~\ref{fig:ALICE_detector}) is the dedicated heavy-ion experiment at the LHC studying properties of the quark-gluon plasma and the QCD phase diagram in general. The detector is designed as a general-purpose heavy-ion detector \cite{ALICE2008} to measure and identify hadrons, leptons, and also photons down to very low transverse momenta.\\

One of ALICE's strengths is the excellent charged particle tracking capability. The tracking is performed in the central barrel, which mainly consists of the Inner Tracking System (ITS) and the Time Projection Chamber (TPC). The ITS \cite{ALICE2010} is a cylindrical six-layered device consisting of three different semiconductor subdetectors: silicon pixel, drift, and strip detectors (SPD, SDD, and SSD). It directly surrounds the beam pipe. Around the ITS, the Time Projection Chamber is placed. The ALICE TPC \cite{ALICE2010b} is mainly filled with neon gas at atmospheric pressure, the gas detector has a radius of $250\mathrm{~cm}$ and a length of $500 \mathrm{~cm}$. Combining the ITS and the TPC, a straight track can be reconstructed within a pseudorapidity interval of $\abs{\eta} < 0.9$.

To measure fully reconstructed jets, also neutral particles have to be detected. For this purpose, a lead-scintillator sampling electromagnetic calorimeter -- the EMCal \cite{ALICE2008b} -- is used. The calorimeter covers the pseudorapidity interval $\abs{\eta} < 0.7$ and roughly $107\degree$ in azimuth. While the EMCal is very sensitive to photons, e.g. from neutral pion decays, the neutron reconstruction capability is less precise.

The VZERO \cite{ALICE2010c} scintillation counter is used for event characterization and centrality estimation.

\section{Jet reconstruction and correction techniques}
\label{sec:Techniques}

Jet reconstruction is usually a multistep procedure. Next to the conceptual definition of jets as hadronized partons, a technical jet definition is necessary. There is no unambiguous jet definition and no common way how to measure a jet. The currently most prevalent jet definition at the LHC is given by the anti-$k_\mathrm{T}$ algorithm \cite{Cacciari2008} implemented in the FastJet \cite{Cacciari2006} package. For all presented analyses, the anti-$k_\mathrm{T}$ algorithm is used to measure signal jets. Additionally, the $k_\mathrm{T}$ algorithm is used for some of the background correction techniques.

The input passed to the jet finding algorithm is given by charged particles and calorimeter clusters. The track cuts are chosen to obtain a uniform charged track distribution in the full $\eta-\phi$ plane.
Additionally, only tracks from $\abs{\eta} < 0.9$ and with $\pT > 150 \mbox{ MeV/}c$ are used in the jet finding procedure.

While \textit{charged jets} are reconstructed using only the charged part of a jet, \textit{full jets} also include neutral jet energy reconstructed from calorimeter clusters. Since the EMCal is also sensitive to charged tracks, the energy of each cluster has to be corrected for charged energy to avoid double-counting. This is done by extrapolating tracks to the EMCal and matching them to the deposited energy. The corrected cluster energy is given by
\begin{equation}
E^\mathrm{corr}_\mathrm{clus} = E^\mathrm{orig}_\mathrm{clus} - \sum{p^\mathrm{matched}}.
\end{equation}
Eventually, only clusters with $E^\mathrm{corr}_\mathrm{clus} > 300$ MeV are passed to FastJet.
After constituent selection, the tracks and calorimeter clusters are clustered by FastJet using the so-called $\pT$-recombination scheme, assuming massless jets.

In ALICE, jets have been measured with several jet resolution parameters (often called \textit{jet radii}) for different collision systems. To avoid edge effects, only jets fully contained within the acceptance are used for further analysis.

Due to the very high multiplicities in heavy-ion collisions, there is a non-negligible contribution of jets at lower $\pT$  made of so-called combinatorial jets mostly containing background. To suppress those jets, a bias on the leading particle's transverse momentum can be imposed. On the other hand, this biases the jets to show a harder fragmentation but, eventually, the systematic uncertainty from the fragmentation bias is smaller than that from combinatorial jets.

\subsection{Background correction}

In principle, every particle that does not originate from the hard parton scattering in question can be considered as background. Like the definition of a jet, also the definition of the background is not unambiguous.
All presented analyses use background correction techniques but the methods differ depending on the considered collision system: While the background subtraction is a large correction to the jet momentum in $\PbPb$ collisions, the background density in $\pp$ and $\pPb$ collisions is roughly smaller by a factor of 100 \cite{ALICE2014c}\cite{Cacciari2008b}.

A typical approach to correct for background is to subtract the background on a jet-by-jet basis. The average background energy density $\rho$ is evaluated using $k_\mathrm{T}$ jets ($\pPb$ and $\PbPb$) or by probing the background in a cone perpendicular to the jet ($\pp$) on an event-by-event basis. Then, the transverse momentum of every jet is corrected for the background momentum density $\rho$ depending on the jet area $A$ by
\begin{equation}
p^\mathrm{corr}_\mathrm{T,\;jet} = p^\mathrm{orig}_\mathrm{T,\;jet} - \rho A.
\end{equation}
In $\pPb$ and $\PbPb$ collisions, region-to-region fluctuations of the background are also taken into account. These are quantified by probing the transverse momentum in randomly distributed cones and comparing it to the average background. Quantitatively, this distribution is given by
\begin{equation}
\delta \pT = \sum_\mathrm{i}{p_\mathrm{T,\,i}-\rho A}, ~~~ A = \pi R^2.
\end{equation}

\subsection{Unfolding}
Spectra in $\pPb$ and $\PbPb$ collisions are corrected for background fluctuations and detector effects -- e.g. from the limited single-particle tracking efficiency -- in an unfolding procedure. In $\pp$ collisions, background fluctuations are neglected and only detector effects are corrected for. Depending on the analysis, Bayesian, SVD, or $\chi^2$-unfolding or also a conceptually simpler bin-by-bin correction is used.
The detector response matrix is created with a full detector simulation using PYTHIA6 to generate jets and GEANT3 for the particle transport through the detector.

\section{Results}
\label{sec:Results}

In the following, the results on jet production measurements are presented in terms of the nuclear modification factors and jet cross section ratios.

The nuclear modification factor shows the effect of the nuclear environment on jet production. It is defined by
\begin{equation}
R_\mathrm{AA} =
\frac
{\left.\frac{\mathrm{d}N}{\mathrm{d}p_\mathrm{T}\mathrm{d}\eta} \right| _\mathrm{AA}}
{\left<T_\mathrm{AA}\right>   \cdot \left.\frac{\mathrm{d}\sigma}{\mathrm{d}p_\mathrm{T}\mathrm{d}\eta} \right| _\mathrm{pp}}
\end{equation}
where $T_\mathrm{AA}$ is the nuclear overlap function \cite{ALICE2013c}, which connects the heavy-ion collisions in $\PbPb$ or $\pPb$ to a reference measurement in pp.

\begin{figure}[h]
\begin{minipage}{0.41\textwidth}
\includegraphics[width=\textwidth]{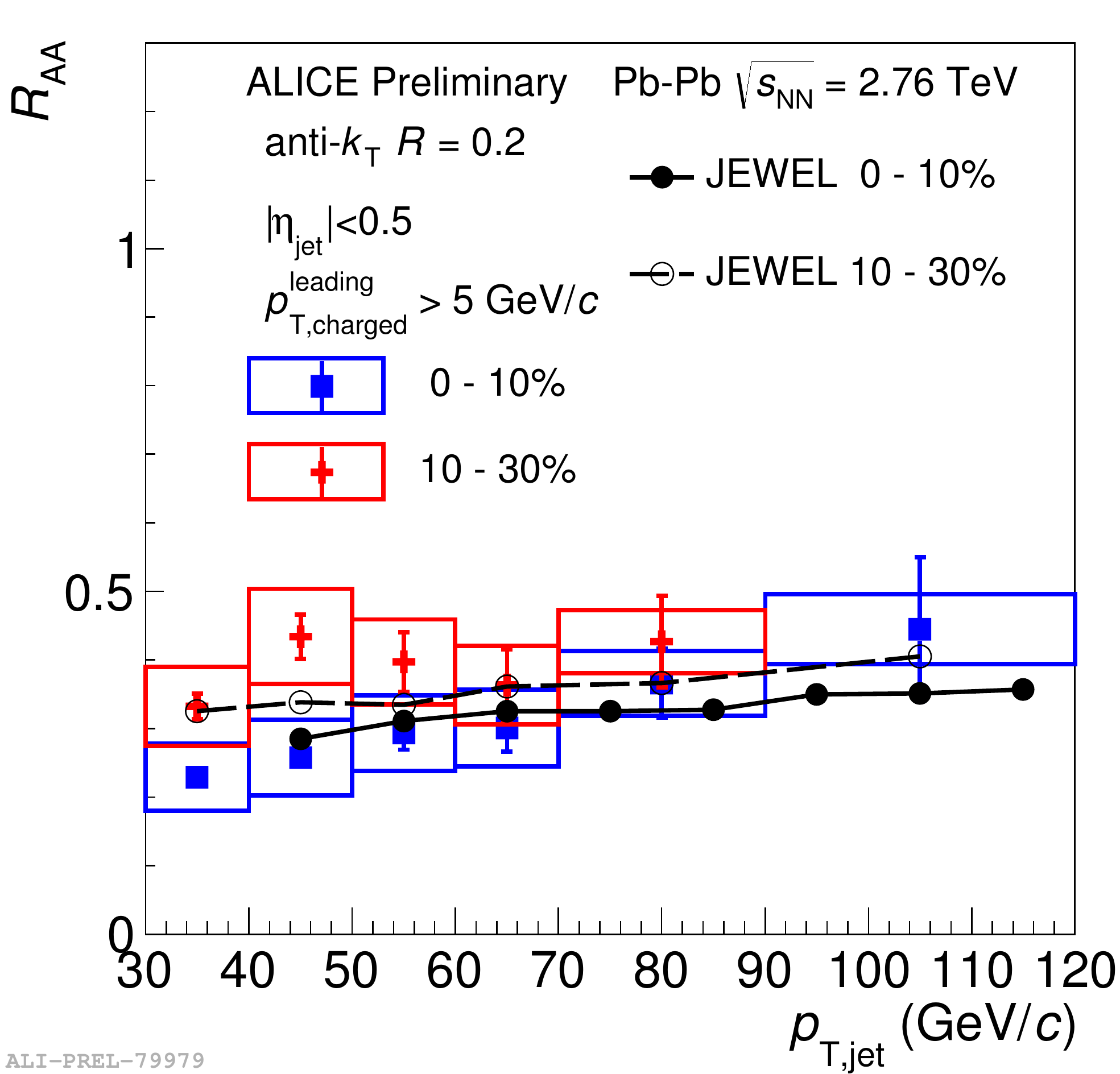}
\caption{\label{fig:PbPb_full_RAA}Nuclear modification factor for $R=0.2$ (Pb--Pb, full jets, $\sqrt{s_\mathrm{NN}} =$ 2.76 TeV)}
\end{minipage} \hspace{2pc}%
\begin{minipage}{0.56\textwidth}
\includegraphics[width=\textwidth]{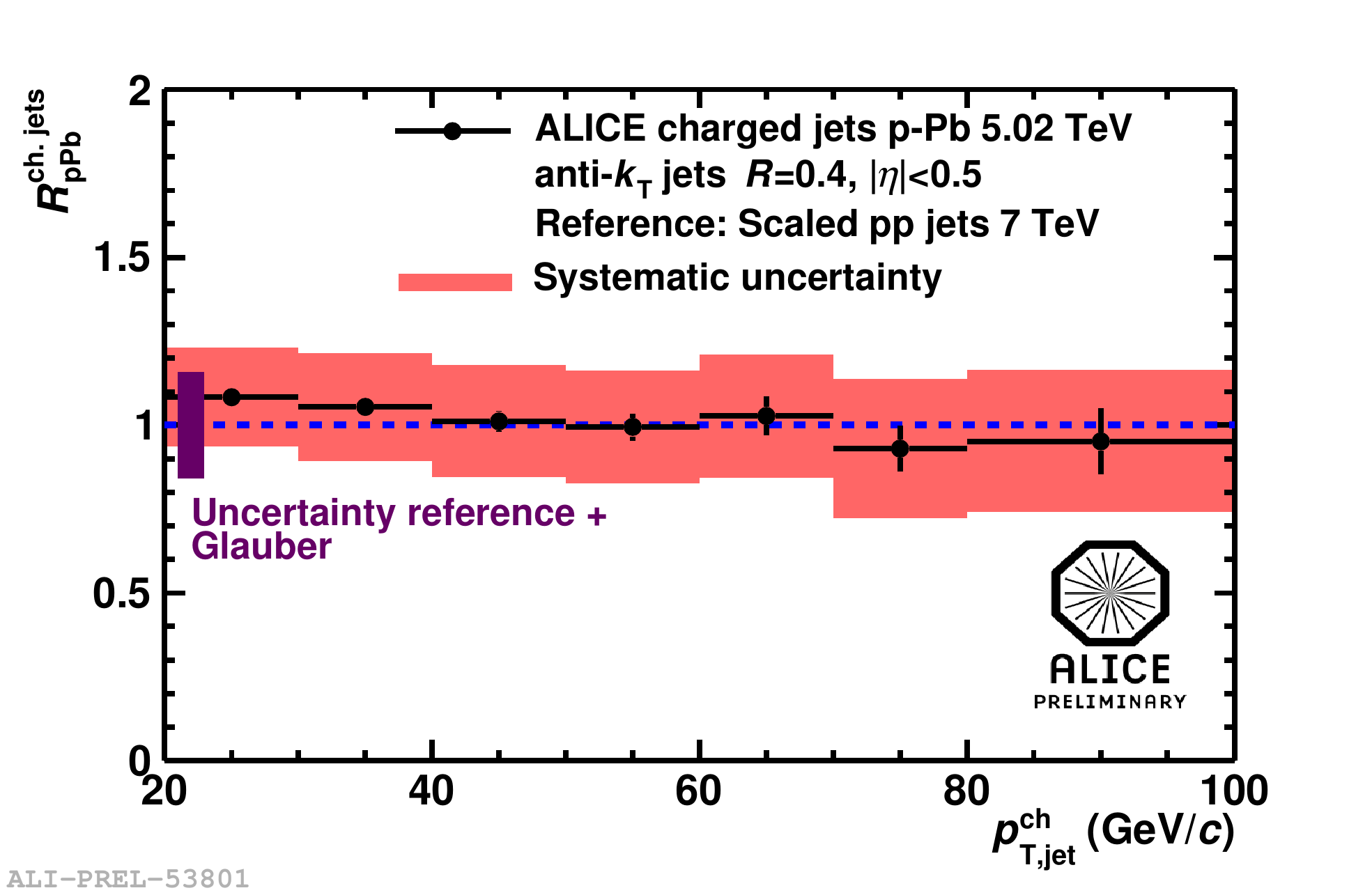}
\caption{\label{fig:pPb_charged_RpPb}Nuclear modification factor for\\ $R=0.4$ (p--Pb, charged jets, $\sqrt{s_\mathrm{NN}} =$ 5.02 TeV)}
\end{minipage} 
\end{figure}

Preliminary ALICE measurements for full jets in $\PbPb$ \cite{Aiola2014} and charged jets in $\pPb$ \cite{Haake2013} are presented here, see Figs.~\ref{fig:PbPb_full_RAA} and \ref{fig:pPb_charged_RpPb}, respectively. The data in $\PbPb$ clearly shows jet quenching, i.e. a strong jet suppression, consistent with the charged hadron result \cite{ALICE2011}. In addition, the measurement shows that the suppression has a centrality dependence and a small transverse momentum dependence.

For $\pPb$ collisions (Fig.~\ref{fig:pPb_charged_RpPb}), no nuclear modification is observed. This is a strong hint that the jet quenching originates mainly from medium effects due to the formation of a quark-gluon plasma and not from cold nuclear effects due to the presence of the Pb-nucleus. 

\begin{figure}[h]
\begin{minipage}{0.345\textwidth}
\includegraphics[width=\textwidth]{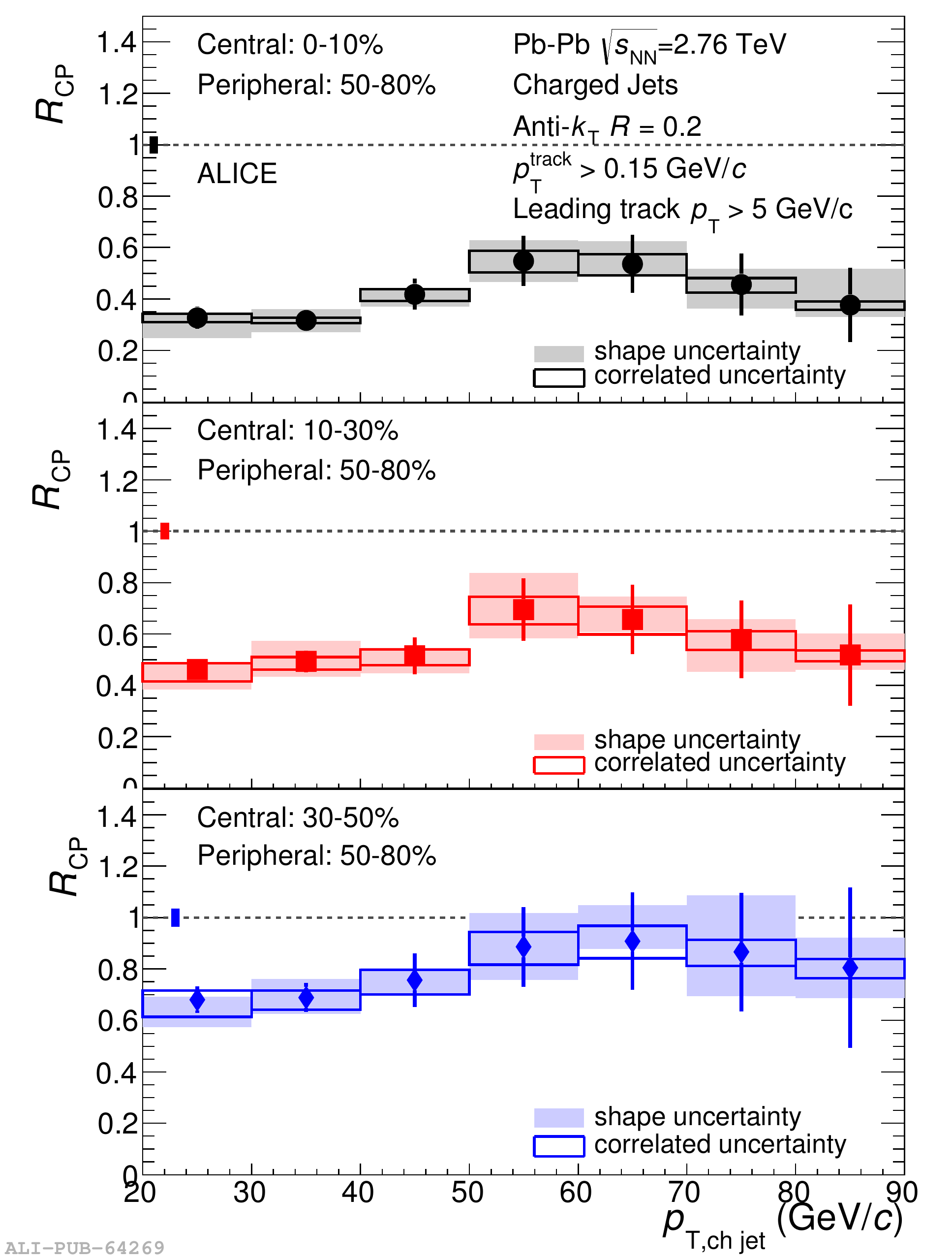}
\caption{\label{fig:PbPb_charged_RCP}Central to peripheral ratio of jets for $R=0.2$ (Pb--Pb, charged jets, $\sqrt{s_\mathrm{NN}} =$ 2.76 TeV)}
\end{minipage}\hspace{1.5pc}%
\begin{minipage}{0.60\textwidth}
\includegraphics[width=\textwidth]{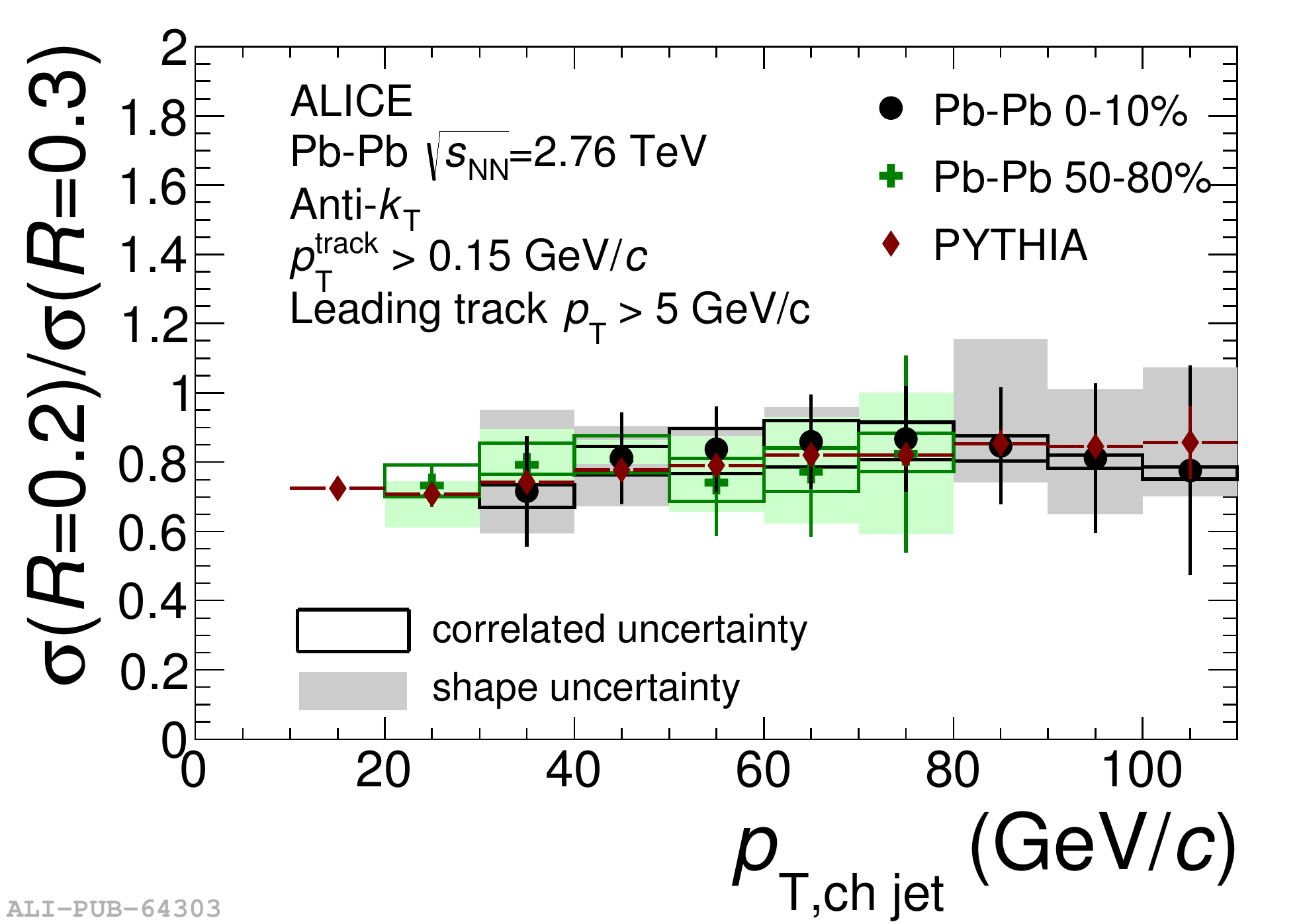}
\caption{\label{fig:PbPb_charged_ratio}Jet cross section ratio compared to PYTHIA calculations (Pb--Pb, charged jets, $\sqrt{s_\mathrm{NN}} =$ 2.76 TeV)}
\end{minipage}
\end{figure}

A conceptually simpler quantity is $R_\mathrm{CP}$, the ratio of central over peripheral jet spectra scaled by the respective number of binary collisions. Like the nuclear modification factor, it is a measure for the nuclear modification but it is simpler because no pp reference is needed. It is shown for charged jets in $\PbPb$ collisions \cite{ALICE2014} in Fig.~\ref{fig:PbPb_charged_RCP}. Like the full jet $R_\mathrm{AA}$, the charged jet $R_\mathrm{CP}$ shows a centrality-dependent suppression that only has a small transverse momentum dependence.

\begin{figure}[h]
\begin{minipage}{0.43\textwidth}
  \centering
  \includegraphics[width=\textwidth]{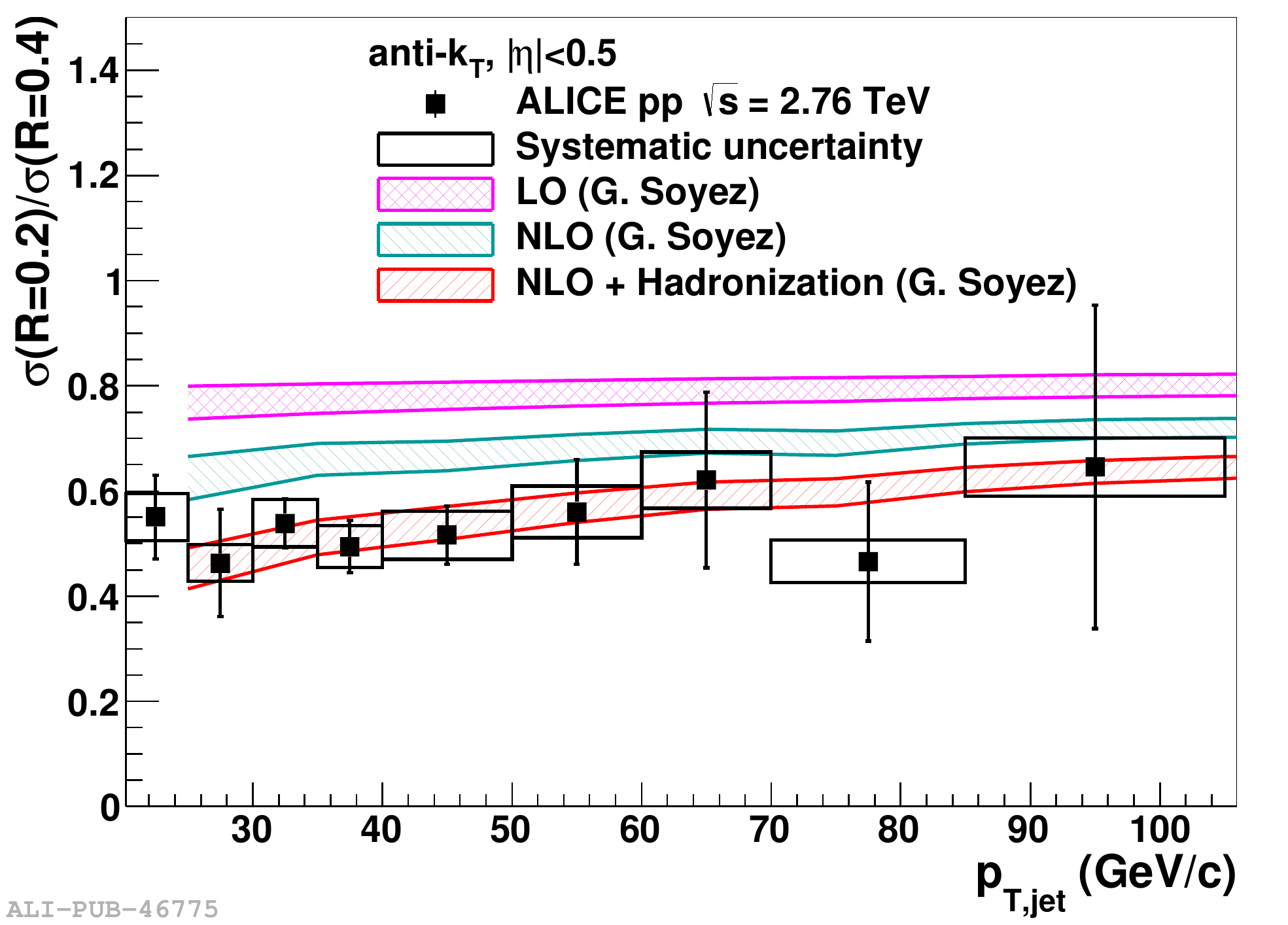}
  \caption{\label{fig:pp_full_ratio}Jet cross section ratio (pp, full jets, $\sqrt{s} =$ 2.76 TeV)}
\end{minipage}\hspace{0.04\textwidth}%
\begin{minipage}{0.47\textwidth}
  \centering
  \includegraphics[width=\textwidth]{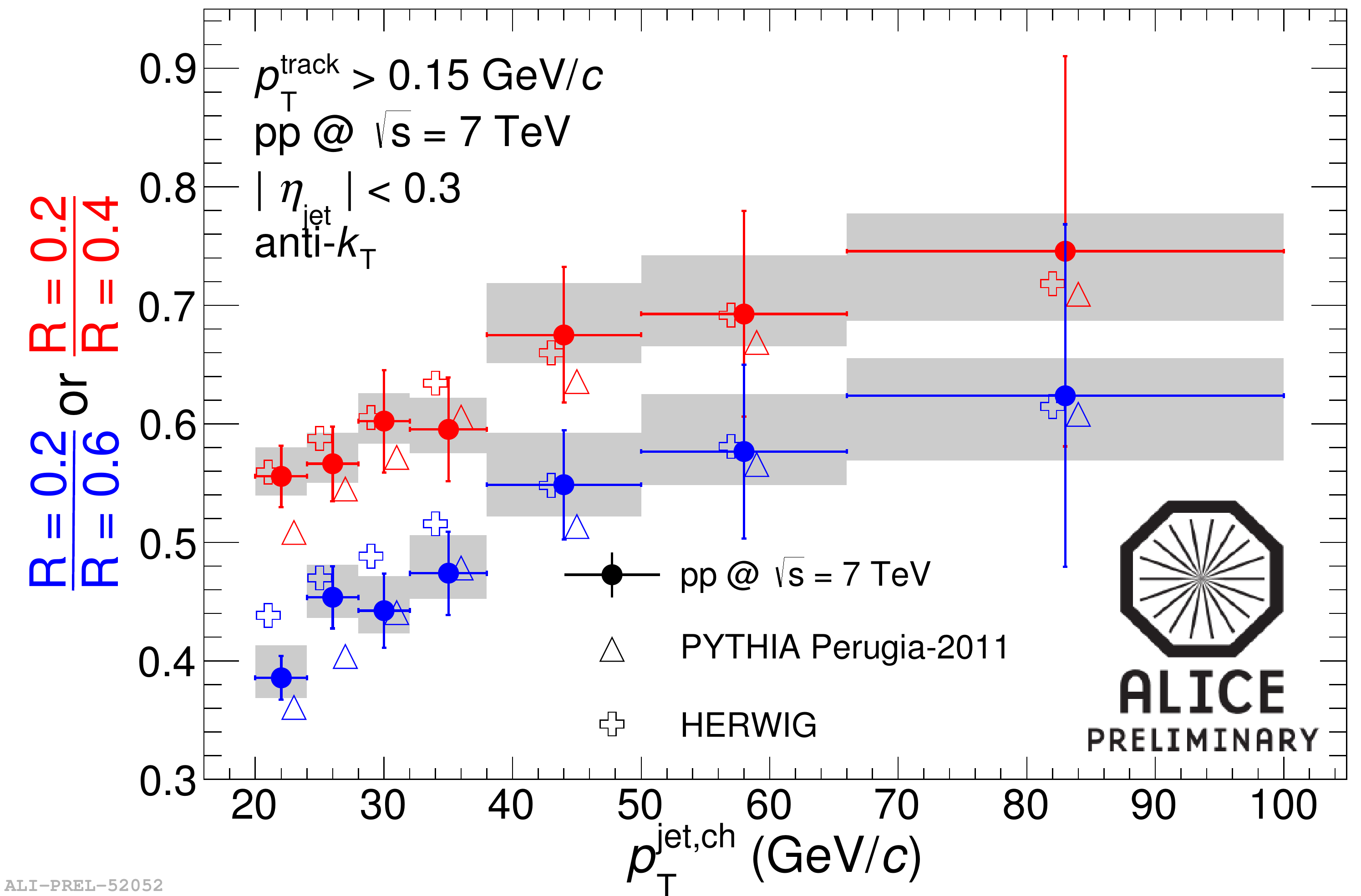}
  \caption{\label{fig:pp_charged_ratio}Jet cross section ratio (pp, charged jets, $\sqrt{s} =$ 7 TeV)}
\end{minipage}
\end{figure}

A basic observable to study the internal radial structure of jets, i.e. jet broadening and narrowing, is the ratio of jet cross sections reconstructed with different resolution parameters.
For $\pp$ collisions, the ratio has been measured for full (Fig.~\ref{fig:pp_full_ratio}) and charged jets (Fig.~\ref{fig:pp_charged_ratio}) at different energies. The full jet analysis in $\pp$ \cite{ALICE2013b} shows a good theory-data agreement when comparing to an NLO calculation with hadronization effects. The charged jet analysis in $\pp$ \cite{Vajzer2013}\cite{ALICE2014b} shows a good agreement with PYTHIA and HERWIG simulations.

\begin{figure}[h]

\begin{minipage}{0.48\textwidth}
  \includegraphics[width=\textwidth]{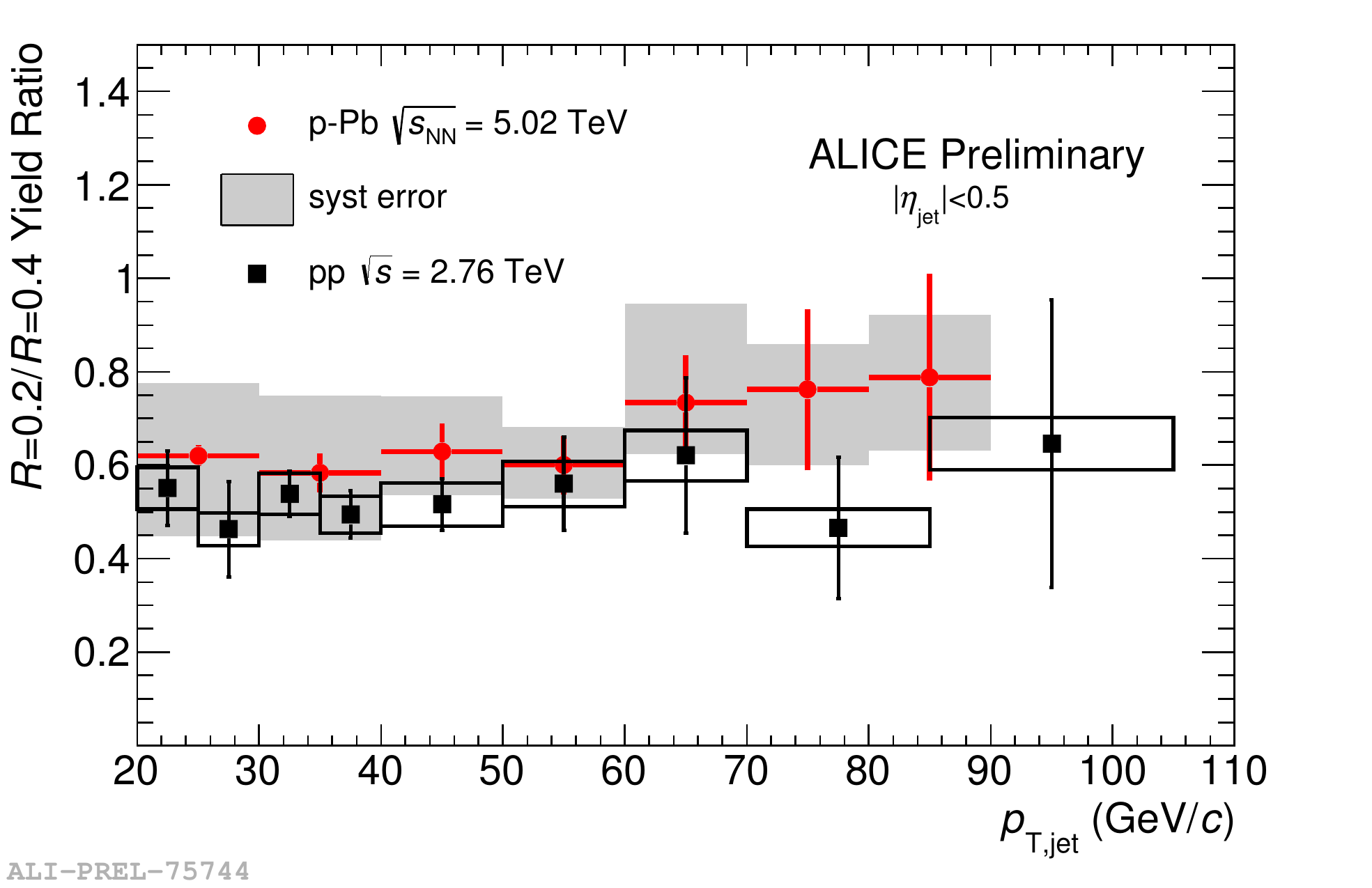}
  \caption{\label{fig:pPb_full_ratio}Jet cross section ratio in $\pPb$ (full jets, $\sqrt{s_\mathrm{NN}} =$ 5.02 TeV) compared to that in pp at $\sqrt{s} =$ 2.76 TeV}
\end{minipage}\hspace{0.04\textwidth}%
\begin{minipage}{0.48\textwidth}
  \includegraphics[width=\textwidth]{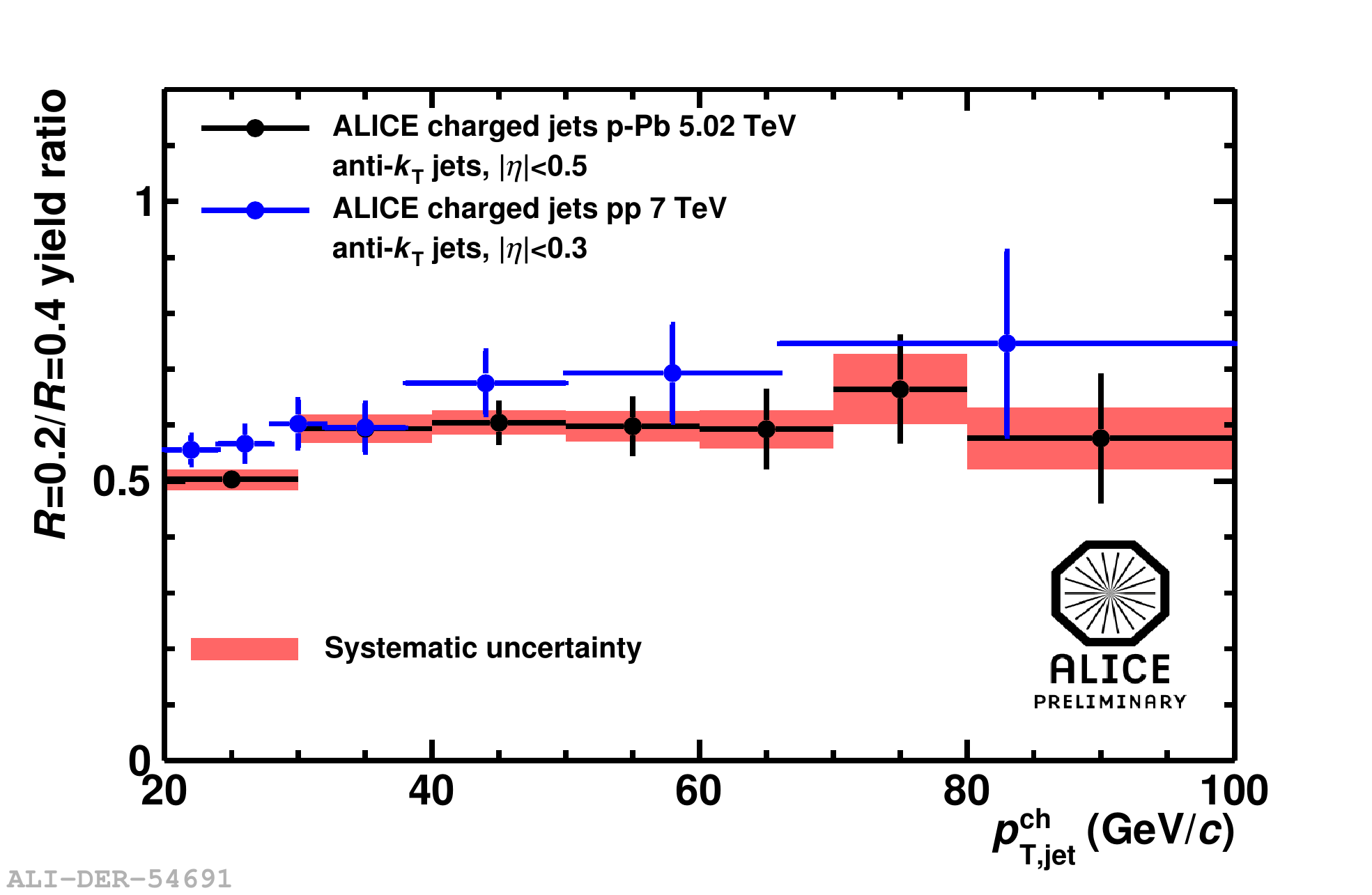}
  \caption{\label{fig:pPb_charged_ratio}Jet cross section ratio in $\pPb$ (charged jets, $\sqrt{s_\mathrm{NN}} =$ 5.02 TeV) compared to that in pp at $\sqrt{s} =$ 7 TeV}
\end{minipage}
\end{figure}

For charged jets in $\PbPb$ (Fig.~\ref{fig:PbPb_charged_ratio}), no significant medium modification of the radial jet structure is observed comparing to $\pp$ collisions. It is compatible with PYTHIA simulations. The same holds for $\pPb$ collisions: The radial structure in both, full (cf. \cite{Connors2014}) and charged jets in $\pPb$, agree well with proton-proton results (Figs.~\ref{fig:pPb_full_ratio} and \ref{fig:pPb_charged_ratio}).

Thus, the radial jet structure is essentially unmodified by the nuclear environment and the medium in $\pPb$ and $\PbPb$ collisions.

\newpage

\section{Summary}
\label{sec:Summary}

ALICE has measured charged and full jets in pp, $\pPb$, and $\PbPb$ collisions at several energies. In pp, the spectra and radial structure ratios for different resolution parameters show a good agreement with (next-to-)leading-order calculations. The same also holds for $\pPb$ collisions: In $\pPb$ collisions, there is no sign for a nuclear modification within the uncertainties. In $\PbPb$ collisions, jet quenching and its centrality dependence has been measured for full and charged jets. Neither in $\pPb$ nor in $\PbPb$, the jet cross section ratio shows a significant modification.\\

\bibliographystyle{iopart-num} %
\bibliography{HPT2014_Proceedings_Haake}

\providecommand{\newblock}{}
\begin{thebibliography}{10}
\expandafter\ifx\csname url\endcsname\relax
  \def\url#1{{\tt #1}}\fi
\expandafter\ifx\csname urlprefix\endcsname\relax\def\urlprefix{URL }\fi
\providecommand{\eprint}[2][]{\url{#2}}

\bibitem{ALICE2008}
Aamodt K {\em et~al.\/} (ALICE Collaboration) 2008 {\em JINST\/} {\bf 3} 8002

\bibitem{ALICE2010}
Aamodt K {\em et~al.\/} (ALICE Collaboration) 2010 {\em JINST\/} {\bf 5} 3003
  (\textit{Preprint} \eprint{1001.0502})

\bibitem{ALICE2010b}
Alme J {\em et~al.\/} (ALICE Collaboration) 2010 {\em Nucl. Instrum. Meth. A\/}
  {\bf 622} 316--367 (\textit{Preprint} \eprint{1001.1950})

\bibitem{ALICE2008b}
Cortese P {\em et~al.\/} (ALICE Collaboration) {\em CERN-LHCC-2008-014\/}

\bibitem{ALICE2010c}
Aamodt K {\em et~al.\/} (ALICE Collaboration) 2010 {\em Phys. Rev. Lett.\/}
  {\bf 105} 252301 (\textit{Preprint} \eprint{1011.3916})

\bibitem{Cacciari2008}
Cacciari M, Salam G~P and Soyez G 2008 {\em JHEP\/} {\bf 04} 063
  (\textit{Preprint} \eprint{0802.1189})

\bibitem{Cacciari2006}
Cacciari M and Salam G~P 2006 {\em Phys. Lett. B\/} {\bf 641} 57--61
  (\textit{Preprint} \eprint{0512210})

\bibitem{ALICE2014c}
Abelev B {\em et~al.\/} (ALICE Collaboration) 2012 {\em JHEP\/} {\bf 03} 053
  (\textit{Preprint} \eprint{1201.2423})

\bibitem{Cacciari2008b}
Cacciari M and Salam G~P 2008 {\em Phys. Lett. B\/} {\bf 659} 119--126
  (\textit{Preprint} \eprint{0707.1378})

\bibitem{ALICE2013c}
Abelev B {\em et~al.\/} (ALICE Collaboration) 2013 {\em Phys. Rev. C\/} {\bf
  88} 44909 (\textit{Preprint} \eprint{1301.4361})

\bibitem{Aiola2014}
Aiola S (ALICE Collaboration) 2014 {\em QM2014 proceedings, to be published\/}
  (\textit{Preprint} \eprint{1408.0479})

\bibitem{Haake2013}
Haake R (ALICE Collaboration) 2013 {\em PoS EPS-HEP2013 (proceedings)\/}  176
  (\textit{Preprint} \eprint{1310.3612})

\bibitem{ALICE2011}
Aamodt K {\em et~al.\/} (ALICE Collaboration) 2011 {\em Phys. Lett. B\/} {\bf
  696} 30--39 (\textit{Preprint} \eprint{1012.1004})

\bibitem{ALICE2014}
Abelev B {\em et~al.\/} (ALICE Collaboration) 2014 {\em JHEP\/} {\bf 03} 013
  (\textit{Preprint} \eprint{1311.0633})

\bibitem{ALICE2013b}
Abelev B {\em et~al.\/} (ALICE Collaboration) 2013 {\em Phys. Lett. B\/} {\bf
  722} 262--272 (\textit{Preprint} \eprint{1301.3475})

\bibitem{Vajzer2013}
Vajzer M (ALICE Collaboration) 2013 {\em PoS EPS-HEP2013 (proceedings)\/}  464
  (\textit{Preprint} \eprint{1311.0148})

\bibitem{ALICE2014b}
Abelev B {\em et~al.\/} (ALICE Collaboration) 2014 {\em to be published\/}

\bibitem{Connors2014}
Connors M (ALICE Collaboration) 2014 {\em LHCP proceedings, to be published\/}
  (\textit{Preprint} \eprint{1409.4655})

\end{thebibliography}

\end{document}